\documentclass[runningheads]{svmult}

\usepackage{makeidx}   % allows index generation
\usepackage{graphicx}  % standard LaTeX graphics tool
\usepackage{subeqnar}  % subnumbers individual equations
\usepackage{multicol}  % used for the two-column index
\usepackage{physprbb}  % modified textarea for proceedings,
\makeindex             % used for the subject index
\begin{document}
\title*{Exclusive Nonleptonic $B$ Decays from QCD Light-Cone Sum Rules}
%
%
%\toctitle{Focusing of a Parallel Beam to Form a Point
%\protect\newline in the Particle Deflection Plane}
% allows explicit linebreak for the table of content
%
%
\titlerunning{Exclusive Nonleptonic $B$ decays from QCD Light-Cone Sum Rules}
% allows abbreviation of title, if the full title is too long
% to fit in the running head
%
\author{Bla\v zenka Meli\'c\inst{1,2}\thanks{Alexander von Humboldt fellow. On the leave of absence from 
Rudjer Bo\v skovi\'c Institute, Zagreb, Croatia.}}
%
%\authorrunning{Bla\v zenka Meli\'c}
% if there are more than two authors,
% please abbreviate author list for running head
%
%
\institute{Institut f\"ur Theoretische Physik und Astrophysik, Julius-Maximilians-Universit\"at W\"urzburg, 
D-97074 W\"urzburg, Germany
\and Institut f\"ur Physik, Johannes-Gutenberg Universit\"at Mainz, D-55099 Mainz, Germany}

\maketitle              % typesets the title of the contribution

\begin{abstract}
We are going to review recent advances in the theory of
exclusive nonleptonic B decays. The emphasis is going to be on the
factorization hypothesis and the role of nonfactorizable contributions
for nonleptonic B decays. In particular, we will discuss more in 
detail calculations of nonfactorizable contributions in the
QCD light-cone sum rule approach and their implications to the
$B \rightarrow \pi \pi$ and $B \rightarrow J/\psi K$ decays.
\end{abstract}

\section{Exclusive Nonleptonic $B$ Decays and Factorization}
Exclusive nonleptonic decays represent a great challenge to theory. They are 
complicated by the hadronization of final states and strong-interaction 
effects between them. Today measurements have already reached sufficient precision to 
examine our knowledge of these effects. In order to make real use of data in 
the determination of fundamental parameters and in testing of the Standard Model, we are forced to 
provide a more accurate estimation of nonperturbative quantities, such as the matrix 
elements of weak operators. 

At the first sight, the nonleptonic $B$ meson decay seems to be simple, 
as far as we essentially consider this decay as a weak decay 
of heavy $b$ quark. We are encouraged to use this argument by the facts that the $b$ quark mass  
is heavy compared to the intrinsic scale of strong interactions and that the $b$ quark 
decays fast enough to 
produce energetic constituents, which separate without interfering with each other. This naive 
picture was supported by the color-transparency argument \cite{Bjorken} 
and 
natural application to nonleptonic two-body decays emerged under the 
name \emph{the naive factorization} (discussed in detail below). However, although predictions from 
the naive factorization are in relatively good agreement with the data (apart from the 
color-suppressed decays), the 
naive factorization provides no insight into the dynamical 
background of exclusive nonleptonic decays. 

The theoretical discussion of the nonleptonic decay starts with the effective weak Hamiltonian, which 
summarizes our knowledge of weak decays at low scales (for a review see \cite{BBL}):
\begin{eqnarray}
{\cal H}_{weak} = \frac{G_F}{\sqrt{2}} \, V_{Qq_1} V_{q_2q_3} \left [ C_1(\mu) {\cal O}_1 +
C_2(\mu) {\cal O}_2 + .....
\right ] \, . 
\label{ham}
\end{eqnarray}
The $V$s represent the Cabibbo-Kobayashi-Maskawa (CKM) matrix elements specified for the 
particular heavy-quark decay $Q \rightarrow q_1 q_2 \overline{q}_3$. 
Strong-interaction effects above some scale $\mu \sim m_b$ are retained in the Wilson coefficients 
$C_i(\mu)$. These coefficients are perturbatively calculable and therefore well known. 
Actually, the weak theory without strong corrections and QED effects 
knows only the operator ${\cal O}_1$, and 
in that case $C_1(M_W) =1$ and $C_2(M_W)=0$. The operator ${\cal O}_2$, defined in 
(\ref{o12}), emerges after taking the gluon exchange into account and 
therefore its contribution is suppressed as $C_2(\mu) \sim \ln(M_W)/\ln(\mu)$. 

The 
main problem persists in the calculation of matrix elements of operators ${\cal O}_i$ in a particular 
process. In (\ref{ham}) we retain only the leading operators ${\cal O}_1$ and 
${\cal O}_2$ and suppress explicitly so called penguin operators, ${\cal O}_{i=3,...,10}$. 
Being multiplied by, in principle, small Wilson coefficients, the penguin operators usually can be neglected 
(except for the penguin-dominated decays), 
but could be extremely important for detection of $CP$ violation in 
$B$ decay \cite{Neubert,Fleischer,Mannel}. 

The four-quark operators ${\cal O}_1$ and ${\cal O}_2$ differ only in their color structure:
\begin{equation}
{\cal O}_1 = (\overline{q}_{1{i}} \Gamma_{\mu} Q_{{i}})
(\overline{q}_{2{j}} \Gamma^{\mu} q_{3{j}}) \, , \qquad\qquad
{\cal O}_2 = (\overline{q}_{1{i}} \Gamma_{\mu} Q_{{j}})
(\overline{q}_{2{j}} \Gamma^{\mu} q_{3{i}}) \, , 
\label{o12}
\end{equation}
where $i$ and $j$ are color indices, and $\Gamma_{\mu}= \gamma_{\mu}(1-\gamma_5)$. 
The color-mismatched operator ${\cal O}_2$ can be 
projected to the color singlet state by using the relation $\delta_{ij} \delta_{kl} = 
1/N_c \, \delta_{il}\delta_{jk} + 2 \, (\lambda^a/2)_{il} \, (\lambda^a/2)_{jk}$, as
\begin{equation}
{\cal O}_2 = \frac{1}{N_c} {\cal O}_1 + 2 \, \tilde{{\cal O}}_1 \, .
\label{proj}
\end{equation}
This projection, as can be seen from (\ref{proj}), results in a relative suppression of the 
${\cal O}_2$ operator 
contribution of the 
order $1/N_c$ ($N_c$ is the number of colors) and in the appearance of the new operator 
$\tilde{{\cal O}}_1$ with the explicit color SU(3) matrices $\lambda^a$:
\begin{equation}
 \tilde{{\cal O}}_1 = (\overline{q}_{1i} \Gamma_{\mu} \frac{\lambda^a}{2} Q_{i})
 (\overline{q}_{2j}\Gamma^{\mu}\frac{\lambda^a}{2} q_{3j}) \, . 
\label{tilde}
\end{equation}
Depending on the process involved, the operators ${\cal O}_1$ and ${\cal O}_2$ can exchange 
their roles, and 
then it is customary to define the effective parameters $a_1$ and $a_2$ as 
\begin{equation}
a_1 = C_1(\mu) + \frac{1}{N_c}C_2(\mu)\, ,  \qquad\qquad\qquad a_2 = C_2(\mu) + \frac{1}{N_c}C_1(\mu) \, . 
\end{equation}
These parameters distinguish between three classes of decay topologies: \\
- \emph{class-1} decay amplitude, where a charged 
meson is directly produced in the weak vertex; 
i.e. in the quark transition $b \rightarrow u d \overline{u}$ with 
${\cal O}_1 = (\overline{d} \Gamma_{\mu} u )(\overline{u}\Gamma^{\mu}b)$:
\begin{equation}
{\cal A}(B \rightarrow \pi^+ \pi^-)  \sim a_1 \langle O_1 \rangle  \, , 
\end{equation}
- \emph{class-2} decay amplitude, where a neutral 
meson is directly produced, i.e. in the quark transition $b \rightarrow c s \overline{c}$ 
with ${\cal O}_2 = (\overline{c} \Gamma_{\mu} c )(\overline{s}\Gamma^{\mu}b)$:
\begin{equation}
{\cal A}(B^+ \rightarrow J/\psi K^+) \sim a_2 \langle O_2 \rangle  \, ,  
\end{equation}
- \emph{class-3} decay amplitude, where both 
cases are possible, but this amplitude is however connected by isospin
symmetry with the class-1 and class-2 decays; i.e. in the quark transition 
$b \rightarrow c s \overline{u}$ with ${\cal O}_1 = (\overline{c} \Gamma_{\mu} u )
(\overline{s}\Gamma^{\mu}b)$:
\begin{equation}
{\cal A}(B^- \rightarrow D^0 K^-) \sim (a_1 + x a_2) \langle O_1 \rangle \, ,   
\end{equation}
where $x$ denotes the nonperturbative factor being equal to one in the flavor-symmetry limit. 

The effective parameters $a_1$ and $a_2$ are defined with respect to the naive 
factorization hypothesis, which assumes that the nonleptonic amplitude can be expressed as 
the product of matrix elements of two hadronic (bilinear) currents, for example:
\begin{equation}
\langle \pi^+ \pi^-|(\overline{d} \Gamma_{\mu} u )(\overline{u}\Gamma^{\mu}b)| B \rangle 
\rightarrow \langle \pi^-|(\overline{d} \Gamma_{\mu} u )|0 \rangle 
\langle \pi^+|(\overline{u}\Gamma^{\mu}b)| B \rangle 
\label{fact}
\end{equation}
and that there is no nonfactorizable exchange of gluons between the $\pi^-$ and the $|\pi^+ B \rangle $ 
system. Effectively, that means that the 'nonfactorizable' matrix element of the $\tilde{{\cal O}}_1$ 
operator (\ref{tilde}), 
is vanishing, due to the projection of the colored current to the physical colorless state.

\subsection{Nonfactorizable Contributions}

The effective parameters $a_1$ and $a_2$ could be generalized 
to parametrize also \emph{the nonfactorizable strong-interaction effects}, for example 
gluon exchanges between 
bilinear currents (i.e. in (\ref{fact})) which introduce nonvanishing contribution from ${\tilde{\cal O}}$ 
operators. Schematically, in the large $N_c$ limit, 
\begin{eqnarray}
a_1 &=& C_1(\mu) + \frac{1}{N_c} C_2(\mu) + 2 C_2(\mu) \xi^{nf}_2(\mu)\, , \nonumber \\
a_2 &=& C_2(\mu) + \frac{1}{N_c} C_1(\mu) + 2 C_1(\mu) \xi^{nf}_1(\mu)\, , 
\label{as}
\end{eqnarray}
where we have explicitly indicated that the nonfactorizable contribution to 
the class-1 and class-2 decays $\xi^{nf}_{i=1,2}$,  
do not necessarily need to be the same, and also they can be process dependent quantities, which will be 
discussed later. 
Theoretically,  
nonfactorizable effects are desirable in order to cancel explicit the $\mu$ dependence of $C_i(\mu)$ and therefore 
of the $a_i$'s. All physical quantities are $\mu$ independent, and because there is no explicit 
$\mu$ dependence of the matrix elements $\langle {\cal O}_i \rangle $ 
multiplying  $C_i(\mu)$, there must be some underlying 
mechanism to cancel the explicit $\mu$ dependence of $a_i$'s persisting in the factorization approach. 
In the calculation of the Wilson coefficients beyond the leading order, also 
the renormalization scheme dependence is presented 
\cite{Buras}. 
Naturally, the parameter $a_2$ is more sensitive on the value of the 
factorization scale and on the renormalization scheme, due to the similar magnitude and 
different sign of the $C_2(\mu)$ and $1/N_c C_1(\mu)$ terms   
(calculated in the NDR scheme and for $\Lambda_{\overline{MS}}^{(5)} = 225 \, GeV$, the 
Wilson coefficients have the following values: $C_1(m_b) = 1.082$ and $C_2(m_b) = -0.185$ 
\cite{BBL}). 
This means also that $a_2$ is more sensitive to any 
additional nonperturbative long-distance contributions. 

The global fit of $a_1$ and $a_2$ parameters 
to the $B$ meson experimental data performed in \cite{NS},  
%with 
%the parametrization $a_1 = C_1(\mu) + \xi C_2(\mu)$ and $a_2 = C_2(\mu) + \xi C_1(\mu)$, 
has shown that the $a_1$ coefficient, being essentially proportional to $C_1(\mu) \sim 1$,
is in the expected theoretical range:
\begin{equation}
a_1 \sim 1.05 \pm 0.10 \, , 
\label{a1}
\end{equation}
while $a_2$ has the fitted value of
\begin{equation}
a_2 \sim 0.25 \pm 0.05 \, . 
\label{a2}
\end{equation}
Compared with the theoretical values calculated with the $C_1$ and $C_2$ stated above, 
we note that 
both fitted values show no explicit indication that there is a significant nonfactorizable contribution 
in $B$ decays. This confirms the 
naive factorization picture, although the simple extrapolation of results in $D$ decays to the $B$ case 
would suggest that the 
$a_2$ coefficient could be negative, meaning a nontrivial cancellation of the $1/N_c$ terms 
and dominance of (negative) $C_2(\mu)$ in (\ref{as}). 
The negative value of $a_2$ in $D$ decays has found its confirmation 
in the large $N_c$ hypothesis of neglecting the higher order $1/N_c$ terms, \cite{BGR},
 and in the QCD sum rule calculation \cite{BS}, where the 
cancellation of the $1/N_c$ part with the explicitly calculated nonfactorizable terms was verified. 

However, there are additional indications that nonfactorizable contributions in $B$ decays cannot be 
simply neglected and deserve to be investigated. 
%In contrast to $a_1$, the parameter $a_2$ exhibits 
%strong factorization scale and renormalization scheme dependence, \cite{Buras}, which in order to be 
%canceled ask for nonfactorizable contributions. 
New experimental data on 
$B$ mesons indicate nonuniversality of the $a_2$ parameter and the strong final-state interaction 
phases in the color-suppressed class-2 decays being proportional to $a_2$ \cite{NP}. 

Therefore, the nonfactorizable contributions must play an important role in nonleptonic 
decays, particularly in the 
color-suppressed class-2 decays, such as the $B \rightarrow J/\psi K$ decay discussed in Sect.4.

\subsection{Models for the Calculation of Nonfactorizable Contributions}

Nowadays, there exist several approaches for the treatment of nonleptonic decays, which try 
to investigate the dynamical background and nonfactorizable 
contributions of such processes. The most exploited ones are 
\emph{the QCD-improved factorization}, \cite{BBNS}, and 
\emph{the PQCD approach} \cite{KLS}. 

\emph{The PQCD approach} claims the perturbativity of the two-body nonfactorizable amplitude if the
Sudakov suppression is implemented into the calculation. The Sudakov form factor suppresses the 
configuration in which the soft gluon exchange could take place, and the amplitude is dominated by 
exchange of hard gluons and therefore perturbatively calculable. 

A somewhat different method is applied in \emph{the QCD-improved factorization}. This method provides 
the factorization formula that separates soft and hard contribution on the basis of large 
$m_b$ expansion.  The leading nonfactorizable strong
interaction can be then studied systematically, while soft (incalculable) 
contributions are suppressed by $\Lambda_{QCD}/m_b$. The method applies to class-1 decays and 
to class-2 decays under the assumption $m_c \ll m_b$. 

None of these models take nonfactorizable soft $O(\Lambda_{QCD}/m_b)$ corrections into account. 
These corrections can be brought under control by using \emph{the light-cone QCD sum rule method} 
\cite{Khodja}. 
This method is going to be discussed more in detail in what follows.

\section{Light-Cone Sum Rules}
All QCD sum rules are based on the general idea of calculating a relevant quark-current correlation 
function and relating it to the hadronic parameters of interest via a dispersion relation. 
Sum rules in hadron physics were already known before QCD was established (for a comprehensive 
introduction to sum rules see i.e. \cite{deRafael}), but have reached wide application in a 
calculation of various hadronic quantities in the 
form of so-called \emph{SVZ sum rules} \cite{SVZ}. The other type of sum rules, \emph{the light-cone 
QCD sum rules} were established for calculation of exclusive amplitudes and form factors 
(\cite{CK} and references therein).  

\subsection{Light-Cone Sum Rules \emph{vs} SVZ Sum Rules}

In order to illustrate application of the QCD sum rules and the main 
differences between SVZ sum rules and light-cone sum rules, we introduce 
an example. 

One of the typical calculation using \emph{the SVZ sum rules} is the estimation of the $B$ meson 
decay constant $f_B$. The starting point is a correlation function defined as
\begin{eqnarray}
F(q^2) = i \int d^4 x e^{i q x}{\langle 0 |} T\{ m_b \overline{u} i \gamma_5 b(x),
m_b \overline{b} i \gamma_5 u(0) \}{| 0 \rangle} \, . 
\end{eqnarray}
In the Euclidean region of $q$ momenta, $q^2 < 0$, we can perform a perturbative calculation 
in terms of quarks and gluons 
by applying the short-distance operator-product expansion (OPE) to the correlation function $F(q^2)$. The 
correlation function is then expressed via a dispersion relation
in terms of the spectral function $\rho^{OPE}$, representing the 
perturbative part, and the quark and gluon 
condensates, i.e. $\langle q \overline{q} \rangle$, $\langle G G \rangle$, etc 
(see for example \cite{Reinders}): 
\begin{equation}
F^{OPE}(q^2) = \frac{1}{\pi} \int_0^{\infty} ds \frac{Im F^{OPE}(s) \, ds }{s-q^2} 
+  A_i(s) \langle O| \Omega_i |0 \rangle \, , 
\end{equation}
where
\begin{equation}
\frac{1}{\pi} Im F^{OPE}(s) = \rho^{OPE}(s) 
\end{equation}
and $A_i$ are perturbative coefficients in front of the vacuum condensates of operators
$\Omega_i = \overline{q}q \, , G G\, , \overline{q} \lambda^a/2 \sigma \cdot G q$ \, , etc. 

On the other hand, in the physical (Minkowskian) region, $q^2 > 0$,
we insert the complete sum over hadronic states 
starting from the ground state $B$ meson, and use a defining relation for $f_B$ : 
$\langle m_b \overline{q} i \gamma_5 b | B \rangle = f_B m_B^2$. The correlation function $F(q^2)$ can then be 
written as 
\begin{equation}
F^{hadron}(q^2) = \frac{m_B^4 f_B^2}{m_b^2-q^2} + \int_{s_0^h}^{\infty} \frac{\rho^{hadron}(s) \,  ds}{s-q^2} \, , 
\end{equation}
where the hadronic spectral density $\rho^{hadron}$ 
contains all higher resonances and non-resonant states with the 
$B$ meson quantum numbers.

By applying \emph{the quark-hadron duality} to these higher hadronic (continuum) states, which means 
assuming that we can replace the continuum of hadronic states, described by the hadronic 
spectral function $\rho^{hadron}(s)$ via a dispersion relation, by the spectral function calculated perturbatively 
in the $q^2 < 0$ region $\rho^{OPE}(s)$,  
we match both sides, $F^{OPE}(q^2) = F^{hadron}(q^2)$, and extract the needed quantity $f_B$.  The replacement is done for 
$s > s_0^B$, where $s_0^B$ is an effective parameter of the order of the mass of the first excited $B$ meson resonance squared. 

In a practical calculation one performs a finite power expansion in $\rho^{OPE}(s)$. 
To improve the convergence of the expansion, 
\emph{the Borel transform} of both sides, $F^{OPE}(q^2)$ and $ F^{hadron}(q^2)$, is considered, 
defined by the following 
limiting procedure
\begin{equation}
{{\cal B} = lim \frac{1}{(n-1)!} (-q^2)^n (\frac{d}{d q^2})^n}
\quad\quad |q^2|, n \rightarrow \infty \, , 
{\frac{|q^2|}{n}} =  {M^2}\;\; fixed \, . 
\end{equation}
$M^2$ is so called Borel parameter. It is determined by the search for stability criteria in a 
sense that, on the one hand, excited and continuum states are suppressed (asks for smaller $M^2$) 
and, on the other hand, the reliable perturbative calculation 
is enabled (asks for larger $M^2$). 

\begin{figure}
\begin{center}
\includegraphics[width=1.0\textwidth]{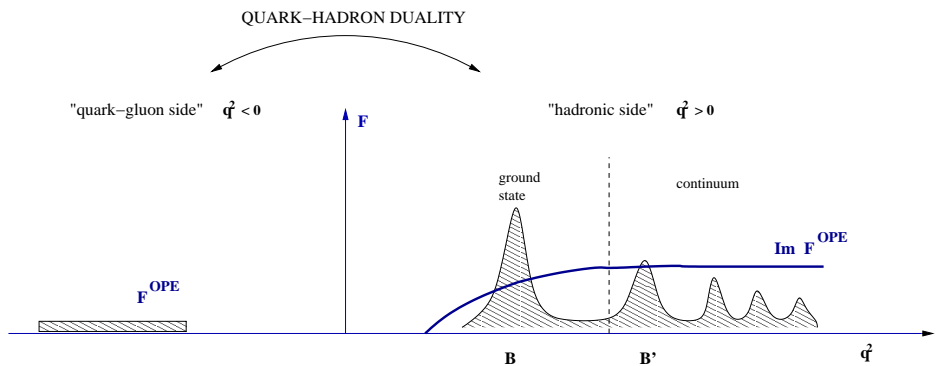}
\end{center}
\caption[]{An illustration of the matching procedure in QCD sum rules}
\label{eps1}
\end{figure}

The general procedure of QCD sum rules is depicted on Fig.~\ref{eps1}. 

For calculating quantities which involve hadron interactions, such as for example the 
$B \rightarrow \pi$ form factor, \emph{the light-cone sum rules} are 
more suitable \cite{Braun}. The correlation function is now defined as a vacuum-to-pion matrix 
element: 
\begin{equation}
F_{\mu} = i \int d^4 x e^{-i p x}
{\langle \pi(q) |} T\{ \overline{u} \gamma_{\mu} b(x),
m_b \overline{b} i \gamma_5 d(0) \} {| 0 \rangle} \, . 
\end{equation}
The calculation follows by performing a light-cone OPE, an expansion 
in terms of the light-cone wave functions of increasing twist (twist = dimension - spin). Physically, it  
means that one performs an expansion in the transverse quark distances in the infinite momentum 
frame, rather than a short-distance 
expansion \cite{Braun}. 
Instead of dealing with the vacuum-to-vacuum quark and gluon condensates (numbers) like in the SVZ sum rules, 
we have now to know the pion distribution amplitude (wave function). The leading twist-2 pion 
distribution amplitude, $\phi_{\pi}$ is defined as 
\begin{equation}
\langle \pi(q) | \overline{u}(x) \gamma_{\mu} \gamma_5 d(0) |0 \rangle = - i q_{\mu} f_{\pi} \int_0^1 
du e^{i u q x} \phi_{\pi}(u) \, . 
\end{equation}
Distribution amplitudes (DAs) describe  distributions of the pion momentum over the pion constituents and 
$u$ denotes the fraction of this momentum, $ 0 < u < 1$
(for a comprehensive paper on the exclusive decays and the light-cone DAs see \cite{Brodsky}). 
The DAs represent a nonperturbative, noncalculable input and their form has to be determined by 
nonperturbative methods and/or somehow extracted from the experiment. 

In the physical region of $(p-q)^2 > 0$  
nothing changes in comparison to the SVZ sum rules. 
We insert the complete set of hadronic states with $B$ meson quantum numbers as before, 
and extract the $B \rightarrow \pi$ form factor from the relation: $\langle \pi(q)|\overline{u}
\gamma_{\mu} b | B (p+q) \rangle = 2 f_{B\pi}^+(p^2) q_{\mu} + ...$. 
The matching procedure follows as described above. 

\section{Nonfactorizable Effects in the Light-Cone Sum Rules}

Although the idea to apply QCD sum rules for calculating nonfactorizable 
contributions in nonleptonic $B$ decays is not the new one, earlier applications were facing 
some problems which have caused unavoidable theoretical uncertainties in their results \cite{Khodja}. 
In the work \cite{Khodja}, a new approach was 
introduced and we are going first to review its main ideas in the 
application to the $B \rightarrow \pi\pi$ decay. 

\subsection{Definitions}
\begin{figure}[t]
\begin{center}
\includegraphics[width=1.0\textwidth]{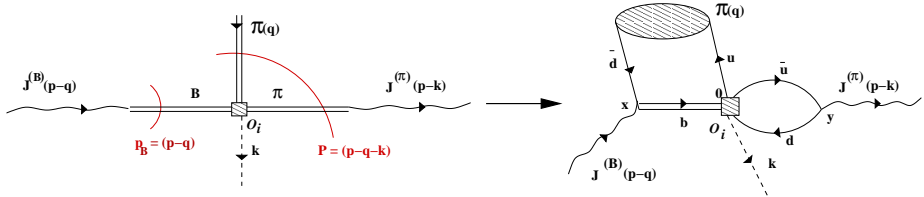}
\end{center}
\caption[]{Sum rule calculation of the $B \rightarrow \pi\pi$ decay. The shaded oval 
region denotes nonperturbative input, 
$\pi$ meson distribution amplitude. The other pion and the $B$ meson are represented by the currents 
$J^{(\pi)}(p-k)$ and $J^{(B)}(p-q)$ respectively. The square represents the four-quark operators ${\cal O}_i$.}
\label{defP}
\end{figure}

The correlator for the $B \rightarrow \pi\pi$ decay given in terms of two interpolating 
currents for the pion and the $B$ meson, $J_{\nu 5}^{(\pi)}= \overline{u} \gamma_{\nu} \gamma_5 d$ and 
$J_5^{(B)} = m_b \overline{b} i\gamma_{5}  d$ respectively, and relevant operators 
${\cal O}_1 = (\overline{d}\Gamma_{\mu}u)(\overline{u}\Gamma^{\mu}b)$ and 
$\tilde{{\cal O}}_1 = (\overline{d}\Gamma_{\mu}\lambda^a/2 u)(\overline{u}\Gamma^{\mu} \lambda^a/2 b)$
looks like:
\begin{equation}
F_{\nu}(p,q,k) = \int d^4 x e^{-i(p-q)x} \int d^4 y e^{i(p-k)y} \langle 0 |
T \{ J_{\nu 5}^{(\pi)}(y) {\cal O}_i(0) J_5^{(B)}(x) \} |\pi(q)  \rangle \, . 
\end{equation}
The transition is defined again between a vacuum and an external pion state. The situation is 
illustrated in Fig.~\ref{defP}. 
One can note an unphysical momentum $k$ coming out from the weak vertex. It was introduced in 
order to avoid the $B$ meson four-momenta before ($p_B = (p-q)$), and 
after ($P$) the decay to be the same, Fig.~\ref{defP}. In such a way, it was prevented that the 
continuum of light states enters the dispersion relation of the $B$ channel. 
States, 
like $D \overline{D}_s^{\ast}$ and $D^{\ast}\overline{D}_s$, have masses smaller 
than the ground state $B$ meson mass and spoil the extraction of the physical $B$ meson. 
These 'parasitic' contributions have caused problems in the earlier application 
of the sum rules \cite{Khodja}. 
There are several other momenta involved into the decay and we take $p^2 = k^2= q^2= 0$ and 
consider region of large spacelike momenta
\begin{equation}
|(p-k)^2| \sim |(p-q)^2| \sim |P^2| \gg \Lambda_{QCD}^2 \, , 
\label{kin}
\end{equation}
where the correlation function is explicitly calculable. 

\begin{figure}
\begin{center}
\includegraphics[width=0.90\textwidth]{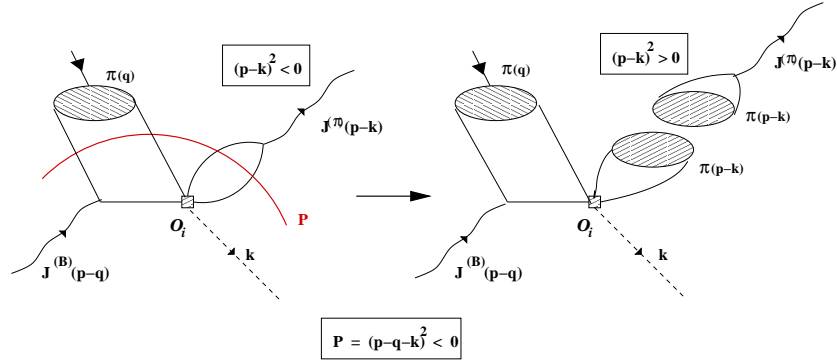}
\vspace*{0.5cm}
\mbox{{(\textbf{a}) Dispersion relation in the pion channel of momentum $(p-k)^2 < 0 $}}
\vspace*{0.5cm}
\includegraphics[width=0.90\textwidth]{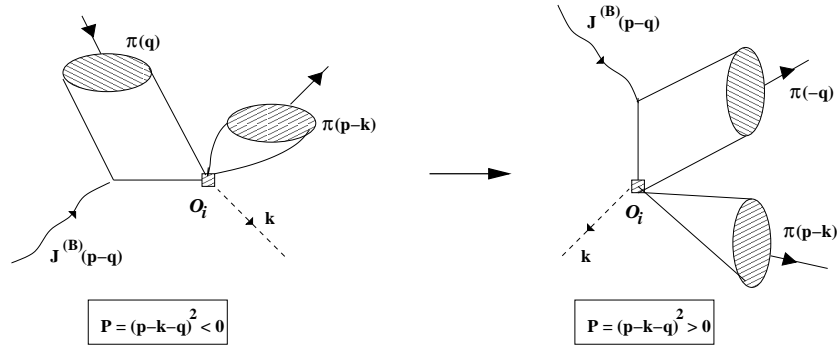}
\vspace*{0.5cm}
\mbox{(\textbf{b}) Analytical continuation of $P^2$ to $P^2 = m_B^2$} 
\vspace*{0.5cm}
\includegraphics[width=0.90\textwidth]{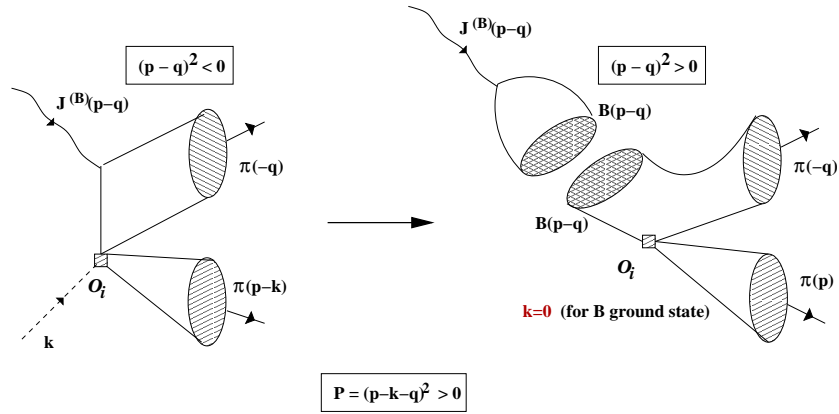}
\vspace*{0.5cm}
\mbox{(\textbf{c}) Dispersion relation in the $B$ meson channel of momentum $(p-q)^2 <0$}
\caption[]{The light-cone sum rule procedure for exclusive decays}
\end{center}
\label{triple}
\end{figure}
\subsection{Procedure}

The procedure which one performs is exhibited in Fig.~3. First, Fig. 3a, one makes a 
dispersion relation in a pion channel of momentum $(p-k)^2$ and applies the quark-hadron duality 
for this channel, as it was explained in Sect.2. Thereafter, to be able later to extract physical $B$ meson state, one has 
to perform an analytical continuation of $P$ momentum to its positive value, $P^2 = m_B^2$. 
This procedure is analogous to the one in the transition from the spacelike to the timelike 
pion form factor, Fig. 3b. Finally, Fig. 3c, a dispersion relation in the $B$ channel of 
momentum $(p-q)^2$ has to be done, together with the application of the quark-hadron duality, 
now in the $B$ channel \cite{Khodja}. 
In such a way we arrive to the double dispersion relation. Apart from somewhat more 
complicated matching procedure, the calculation  otherwise follows in a standard way. 
\subsection{Results and Implications in the $B \rightarrow \pi\pi$ decay}
In \cite{Khodja}, first, the factorization of the ${\cal O}_1$ operator contribution in 
the $B \rightarrow \pi\pi$ decay was confirmed. 
The  soft nonfactorizable contributions due to the $\tilde{\cal O}_1$ operator, which 
express the exchange of soft gluons between two pions in Fig. \ref{defP} 
were then calculated. Nonfactorizable soft contributions appear from the 
absorption of a soft gluon emerging from the light-quark loop 
$\overline{u}d$ in Fig. \ref{defP}, 
by the distribution amplitude of the outcoming pion $\pi^+(-q)$ and there are of the 
higher, twist-3 and twist-4 order in comparison to the factorizable contributions. 

Nonfactorizable 
soft corrections appeared to be numerically small ($\sim 1\%$) and suppressed by $1/m_b$. 
Therefore, their impact on the complete decay amplitude was shown to be of the same 
order as that of hard nonfactorizable contributions calculated in the QCD-improved factorization 
approach \cite{BBNS}. Also, the calculation has shown no imaginary phase from the soft 
contributions, whereas aforementioned hard nonfactorizable contributions get small complex phase because of 
the final state rescattering due to the hard gluon exchange.

\section{Nonfactorizable Effects for $B \rightarrow J/\psi K$}

The $B \rightarrow J/\psi K$ decay was considered in \cite{mi}. As it was emphasized at the 
beginning, this decay belongs to the color-suppressed class-2 decays in which one expects large nonfactorizable 
contributions. The confirmation of this assumption seems to be also found experimentally. Namely, 
there is a discrepancy between the experiment and the naive factorization prediction by at least a 
factor of 3 in the branching ratio.  
The Hamiltonian which describes the decay is given as
\begin{equation}
H_W = \frac{G_F}{\sqrt{2}} V_{c b} V_{c s}^*
\left[ (C_2(\mu) + \frac{1}{N_c} C_1(\mu)) {\cal O}_2  + 2\, C_1(\mu) \tilde{{\cal O}}_2 \right] \, , 
\label{eq:ham0}
\end{equation}
with the operators ${\cal O}_2 = (\overline{c} \Gamma_{\mu} c)(\overline{s} \Gamma^{\mu} b)$ 
and $\tilde{{\cal O}}_2 = (\overline{c} \Gamma_{\mu} \frac{\lambda_a}{2} c)
(\overline{s} \Gamma^{\mu} \frac{\lambda_a}{2} b)$. 
In the factorization approach, 
the matrix element of $\tilde{{\cal O}}_2$ vanishes, and the factorized matrix element of the operator 
${\cal O}_2$ is given by
\begin{eqnarray}
\langle  J/\psi(p) K(q) |  {\cal O}_2 | B(p+q) \rangle &=& \langle J/\psi(p) | \overline{c} \Gamma_{\mu} c
| 0 \rangle
\langle K(q)| \overline{s} \Gamma^{\mu} b | B(p+q) \rangle \nonumber \\
&=& 2 \epsilon \cdot q \, m_{J/\psi}  f_{J/\psi} F^+_{BK}(m_{J/\psi}^2) \, . 
\label{eq:fac}
\end{eqnarray}
$F^+_{BK}(m_{J/\psi}^2)$ is the $B \rightarrow K$ transition form factor 
calculated using the light-cone sum rules, in a way 
enlightened in Sect.2.1 on the example of $B \rightarrow \pi$ form factor calculation, 
and $f_{J/\psi}$ is the $J/\psi$ decay constant. By evaluating numerically the $B \rightarrow J/\psi K$ 
branching ratio with the NLO Wilson coefficients used in Sec.1.1 
and with the numerical input taken from \cite{mi}, we arrive to 
\begin{equation}
{\cal B}(B \rightarrow J/\psi K)^{fact} = 3.3 \cdot 10^{-4}\,.
\label{eq:BRnf}
\end{equation}
This has to be compared with the recent measurements \cite{exp}
\begin{eqnarray}
{\cal B}(B^+ \rightarrow J/\psi K^+) &=& (10.1 \pm 0.3 \pm 0.5) \cdot 10^{-4} \, , 
\nonumber \\
{\cal B}(B^0 \rightarrow J/\psi K^0) &=& (8.3 \pm 0.4 \pm 0.5)\cdot 10^{-4} \, . 
\label{eq:BRexp}
\end{eqnarray}
It is clear that there a discrepancy between the naive factorization prediction and the experiment. 
\begin{figure}[b]
\begin{center}
\includegraphics[width=0.75\textwidth]{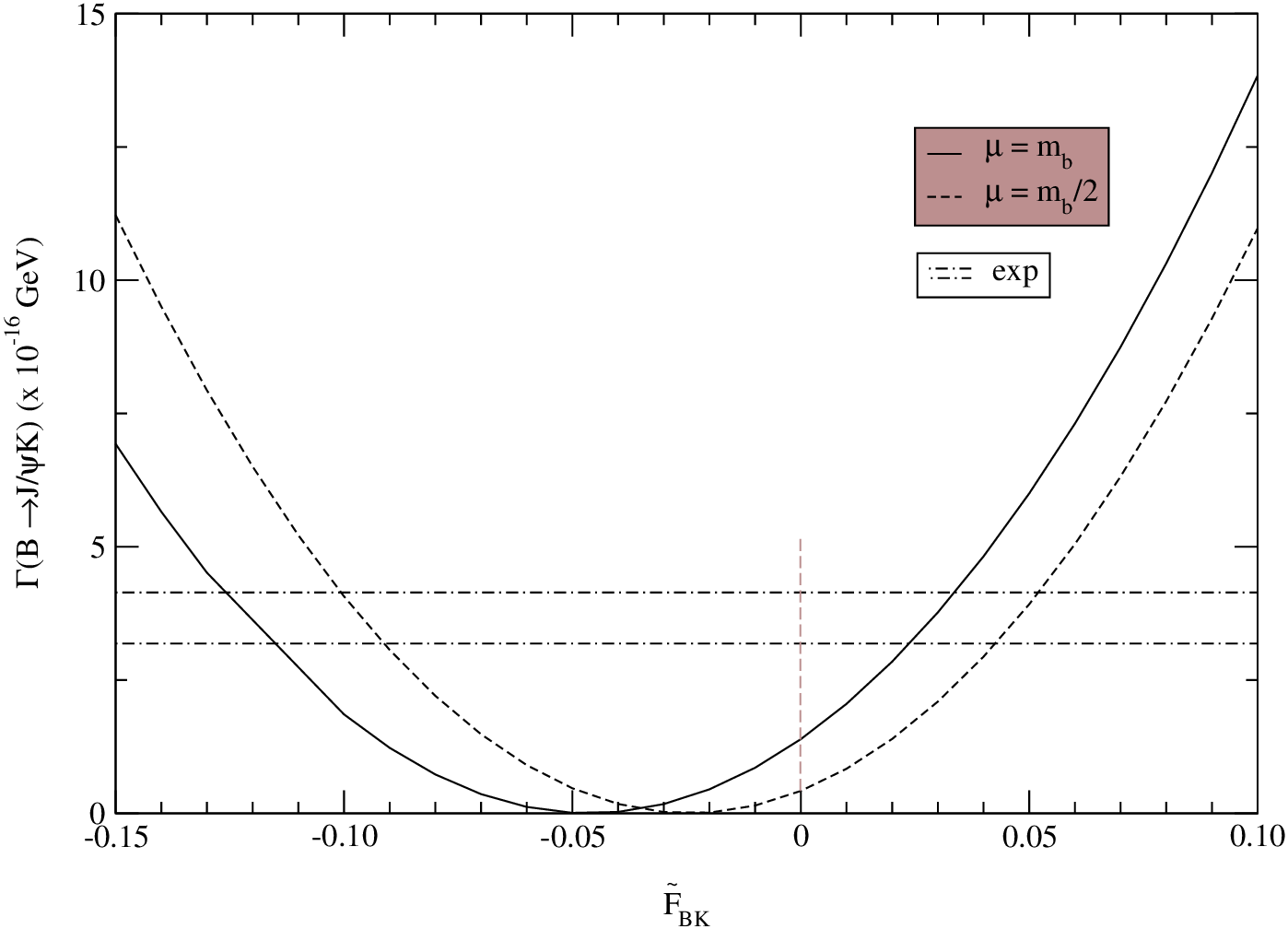}
\end{center}
\caption[]{The partial width for $B \rightarrow J/\psi K$ as a function
of the nonfactorizable amplitude $\tilde{F}_{BK}$. The dashed-dotted lines denote 
the experimental region.}
\label{gamma}
\end{figure}

To be able to discuss the impact of the nonfactorizable term ${\tilde{\cal O}}_2$, we 
parametrize the $\langle J/\psi K | H_W | B \rangle$ amplitude in terms of the $a_2$ parameter as
\begin{equation}
\langle J/\psi K | H_W | B \rangle = \sqrt{2}\, G_F \, V_{c b} V_{c s}^* \, \epsilon \cdot q \,
m_{J/\psi} f_{J/\psi} F_{BK}^+(m_{J/\psi}^2)\, a_2 \, , 
\label{result}
\end{equation}
where
\begin{equation}
a_2 = C_2(\mu) + \frac{C_1(\mu)}{3} +
2 C_1(\mu) \frac{\tilde{F}_{BK}^+(\mu)}{F_{BK}^+(m_{J/\psi}^2)} \, . 
\label{eq:a2def}
\end{equation}
The part proportional to $\tilde{F}_{BK}^+$ represents the contribution from
the ${\tilde{\cal O}}_2$ operator
\begin{equation}
\langle J/\psi K | \tilde{\cal O}_2(\mu) | B \rangle = 2 \epsilon \cdot q \,
m_{J/\psi} f_{J/\psi} \tilde{F}_{BK}^+(\mu^2)
\end{equation}
and  $\tilde{F}_{BK}^+ =0$ corresponds to the naive factorization result,
Eq. (\ref{eq:fac}).

By using the parametrization (\ref{result}) we can extract the $a_2$ coefficient from experiments 
(\ref{eq:BRexp}). The measurements yield 
\begin{equation}
|a_2^{exp}| = 0.29 \pm 0.03 \, . 
\label{eq:a2exp}
\end{equation}

On the other hand, the naive factorization with the NLO Wilson coefficients \cite{Buras} produces
\begin{equation}
a_{2, \, NLO}^{fact} = 0.176\,  |_{\mu \simeq m_b}\, . 
\label{eq:a2nf}
\end{equation}
The value (\ref{eq:a2nf}) is 
significantly below the value extracted from the experiment, 
although one should not forget a strong $\mu$ dependence of $a_2^{fact}$.

In Fig.~\ref{gamma} we show the partial width for $B \rightarrow J/\psi K$ as a function
of the nonfactorizable amplitude $\tilde{F}_{B \rightarrow K}$.
The zero value of $\tilde{F}_{BK}$ corresponds to the factorizable prediction.
There exist
two ways to satisfy the experimental demands on $\tilde{F}_{BK}$. Following the
large $1/N_c$ rule \cite{BGR}, one can argue that there is a cancellation between
$1/N_c$ piece of the factorizable part and the nonfactorizable contribution 
(\ref{eq:a2def}). This would ask for the relatively small and negative value of $\tilde{F}_{BK}$. The
other possibility is to have even smaller, but positive values for $\tilde{F}_{BK}$, which then
compensate the overall smallness of the factorizable part and bring the theoretical
estimation for $a_2$ in accordance with experiment.

One can note significant
$\mu$ dependence of the theoretical expectation for the partial width in Fig.~\ref{gamma}, which
brings an uncertainty in the prediction for $\tilde{F}_{BK}(\mu)$
in the order of  $30 \%$. This uncertainty is
even more pronounced for the positive solutions of $\tilde{F}_{BK}(\mu)$.
The values for $\tilde{F}_{BK}^+$ extracted from experiments
\begin{eqnarray}
\tilde{F}_{BK}^+(m_b) = 0.028 \qquad {\rm or} \qquad \tilde{F}_{BK}^+(m_b)= -0.120 \, ,
\\
\tilde{F}_{BK}^+(m_b/2) = 0.046 \qquad {\rm or} \qquad \tilde{F}_{BK}^+(m_b/2)= -0.095 \, .
\label{eq:fexp2}
\end{eqnarray}
%which gives a two-fold ambiguity of the result, reflecting the unknown sign of the $a_2$.
clearly illustrate the $\mu$ sensitivity of the nonfactorizable part.

In what follows we calculate the nonfactorizable contribution $\tilde{F}_{BK}^+$ 
which appears due to the exchange of soft gluons using the
QCD light-cone sum rule method.
\begin{figure}[b]
\begin{center}
\includegraphics[width=1.0\textwidth]{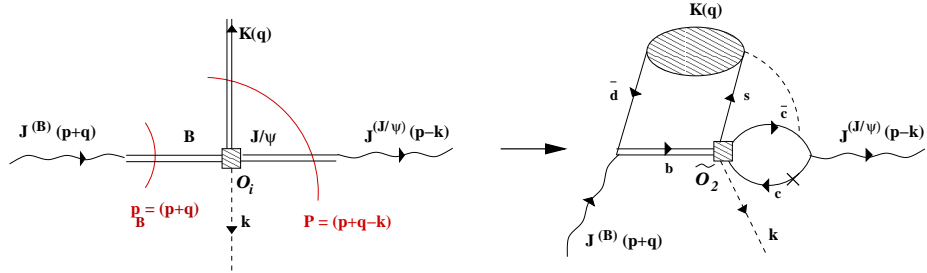}
\end{center}
\caption[]{Sum rule calculation of the $B \rightarrow J/\psi K$ decay. The dashed line denote an exchange 
of a soft gluon for ${\cal O}_i = \tilde{\cal O}_2$ 
and the cross stands for other possible attachment of a soft gluon.}
\label{defJpsi}
\end{figure}
\subsection{Light-Cone Sum Rule Calculation}
The light-cone sum rule calculation starts by considering the correlator
\begin{equation}
F_{\nu}(p,q,k) =
i^2 \int d^4 x e^{-i(p+q)x} \int d^4 y e^{i(p-k)y} \langle K(q) |
T \{ J_{\nu}^{(J/\psi)}(y) {\cal O}(0) J_5^{(B)}(x) \} | 0 \rangle
\end{equation}
with the interpolating currents $J_{\nu}^{(J/\Psi)} =
\overline{c} \gamma_{\nu} c$
and $J_5^{(B)} = m_b \overline{b} i \gamma_5 u$. The kinematics is the same 
as defined above in (\ref{kin}), with the 
exception that now $p^2 = m_{J/\psi}^2$. More explicitly the configuration is shown in 
Fig. \ref{defJpsi}. 

The estimation of nonfactorizable contributions was performed for the exchange of soft gluons (shown by the dashed line 
in Fig. \ref{defJpsi}) and follows essentially steps of derivation explained in Sect. 3.2 for 
$B \rightarrow \pi\pi$ decay. Nonfactorizable contribution of the ${\cal O}_1$ operator appears first 
at $O(\alpha_s^2)$. Nonvanishing result at the one-gluon level 
includes contribution of the $\tilde{\cal O}_2$ operator and the leading contributions are given in terms of 
twist-3 and twist-4 kaon distribution amplitudes which contribute in the same order. 
Technical peculiarities of the calculation can be found in \cite{mi}. 

\subsection{Results and Implications}

The results can be summarized as follows. Soft nonfactorizable twist-3 and twist-4 contributions, expressed in terms of 
$\tilde{F}^+_{BK}$ are 
$\tilde{F}^+_{BK,tw3}(\mu_b) = 0.003 - 0.0055$ and $\tilde{F}^+_{BK,tw4}(\mu_b) = 0.006 - 0.0012$
and the final value is 
\begin{equation}
\tilde{F}^+_{BK}(\mu_b) = 0.009 - 0.0017 \, . 
\label{eq:fth}
\end{equation}
where $\mu_b = \sqrt{M_B^2 -m_b^2} \simeq m_b/2$. 
The wide range prediction for $\tilde{F}^+_{BK}$ appears due to the variation of sum rule parameters. 

First, we note that the nonfactorizable contribution (\ref{eq:fth}) is much 
smaller than the $B \rightarrow K$ transition form factor $F_{BK}^+ = 0.55 \pm 0.05$, which 
enters the factorization prediction (26). It is also significantly 
smaller than the value (\ref{eq:fexp2}) extracted from experiments. Nevertheless, its influence on the
final prediction for $a_2$ is significant, because of the large coefficient $2 C_1$ multiplying
it. Further, one has to emphasize that $\tilde{F}_{BK}^+$ is a positive 
quantity. Therefore, we do not find a theoretical support for the large $N_c$ limit 
assumption discussed in Sect.4.1, that the factorizable part proportional to $C_1(\mu)/3$
should at least be partially cancelled by the nonfactorizable part. Our result also
contradicts the result of the earlier application of QCD sum rules to 
$B \rightarrow J/\psi K$ \cite{KR3}, where negative and somewhat larger value for
$\tilde{F}_{BK}^+$ was found. However, earlier applications  
of QCD sum rules to 
exclusive $B$ decays exhibit some deficiencies discussed in \cite{Khodja}.

Using the same values for the NLO Wilson coefficients as in Sect.2, one gets from (\ref{eq:fth})
the following value for the effective coefficient $a_2$:
\begin{equation}
a_2 \sim 0.14 -0.17\, |_{\mu = \mu_b} \, .
\label{eq:a2calc}
\end{equation}

Although the soft correction contributes in the order of $\sim 30 \% - 70\%$, the net result
(\ref{eq:a2calc}) is still by
approximately factor of two smaller than the experimentally determined value (\ref{eq:a2exp}).

We would like to discuss our results for soft nonfactorizable contributions in
comparison with the hard
nonfactorizable effects calculated in the QCD-improved factorization approach. 
The best thing would be to calculate both soft and hard contributions inside the same model. In principle,
the light-cone sum rule approach presented here enables such a calculation, although the estimation
of hard nonfactorizable contributions is technically very demanding, involving a 
calculation of two-loop diagrams.  Therefore, we proceed with the QCD-improved
factorization estimations for the hard nonfactorizable contributions.

After including the hard nonfactorizable corrections, the $a_2$ parameter 
(\ref{eq:a2calc}) is as follows
\begin{equation}
a_2 =  \left [ C_2(\mu) + \frac{C_1(\mu)}{3} + 2 C_1(\mu) \left ( \alpha_s F(\mu)^{hard} +
\frac{\tilde{F}^+_{BK}(\mu)}{F^+_{BK}} \right ) \right ] \, . 
\label{a2corr} 
\end{equation}

The estimations done in the QCD-improved factorization \cite{Cheng} show hard-gluon
exchange corrections to the naive factorization result in the order of $\sim 25\%$, 
predicted by the LO calculation with the twist-2 kaon distribution
amplitude. Unlikely large corrections are obtained by the inclusion
of the twist-3 kaon distribution amplitude. Anyhow,
due to the obvious dominance of soft contributions to the twist-3
part of the hard corrections in the BBNS approach \cite{BBNS},
it is very likely that some double counting of soft
effects could appear if we naively compare the results. 
Therefore, taking only the twist-2 hard nonfactorizable corrections from \cite{Cheng} into
account, recalculated at the $\mu_b$ scale, our  prediction (\ref{eq:a2calc}) changes to
\begin{equation}
a_2 = 0.16 - 0.19 \, |_{\mu = \mu_b}
\label{eq:a2final}
\end{equation}
The prediction still remains too small to explain the data.

Nevertheless, there are several things which have to be
stressed here in connection with the result. 
Soft nonfactorizable contributions are at least equally important as
nonfactorizable contributions from the hard-gluon exchange, and can be even 
dominant. Soft nonfactorizable
contributions are positive, and the same seems to be valid for 
hard corrections. While 
hard corrections have an imaginary part, in the
soft contributions the annihilation and the penguin topologies as potential sources
for the appearance of an imaginary part were not discussed. 
A comparison between the result (\ref{eq:a2final}) and the 
experimental value $|a_2| \sim 0.3$ for $B \rightarrow J/\psi K$ decay,  
with the recently deduced $a_2$ parameter from the color-suppressed 
$\overline{B}^0 \rightarrow D^{(\ast)0} \pi^0$ decays, $|a_2| \sim 0.4 - 0.5$ \cite{NP}, 
provides clear evidence for the nonuniversality of the $a_2$ parameter in color-suppressed decays. 

\section{Conclusions}

We have reviewed recent progress in the understanding of the underlaying dynamics of  
exclusive nonleptonic decays, with the emphasis on the nonfactorizable corrections to the 
naive factorization approach. In the calculation of nonfactorizable contributions, 
we have focused to QCD light-cone sum rule 
approach 
and have shown results for 
$B \rightarrow \pi\pi$ \cite{Khodja} and $B \rightarrow J/\psi K$ \cite{mi} decays. 

The QCD-improved factorization 
method is reviewed in this volume by M. Neubert, \cite{Neubert}. 

\section*{Acknowledgment}

I would like to thank R. R\"uckl for a collaboration on the subjects discussed in this 
lecture and A. Khodjamirian for numerous fruitful
discussions and comments. The support by the Alexander von Humboldt Foundation is 
gratefully acknowledged. The work was also partially supported 
by the Ministry of Science and Technology 
of the Republic of Croatia under Contract No. 0098002.

%\appendix
%
%\section*{Appendix}
%

%INDEX%%%%%%%%%%%%%%%%%%%%%%%%%%%%%%%%%%%%%%%%%%%%%%%%%%%%%%%%%%%%%%%
% Please check with the editor of your book whether he plans to
% include a "mutual" subject index - if so, please code your entries
% in the standard syntax. For your own purposes you may print your
% "personal" index by using the following commands:
%
%\clearpage
%\addcontentsline{toc}{section}{Index}
%\flushbottom
%\printindex
%%%%%%%%%%%%%%%%%%%%%%%%%%%%%%%%%%%%%%%%%%%%%%%%%%%%%%%%%%%%%%%%%%%%%

\end{document}